# SUCCESSFUL SUPERSYMMETRIC INFLATION

SUBIR SARKAR

*Theoretical Physics, University of Oxford, 1 Keble Road, Oxford OX1 3NP, U.K*

The temperature fluctuations in the cosmic microwave background observed by COBE provide strong support for an inflationary phase in the early universe, below the GUT scale. We argue that a singlet field in a hidden sector of an effective supergravity theory yields the required inflationary potential without fine tuning. Reheating occurs to a temperature low enough to avoid the gravitino problem, but high enough to allow subsequent baryogenesis. Two observational consequences are that gravitational waves contribute negligibly to the microwave background anisotropy, and the spectrum of scalar density perturbations is 'tilted', improving the fit to large-scale structure in an universe dominated by cold dark matter.

## 1 Introduction

Although inflation [1] is an attractive solution to the horizon/flatness problems of the standard Big Bang model and to the cosmological monopole problem of GUTs, it has yet to find a compelling physical basis.[2] Interest in this question has been rekindled by the COBE [3] discovery of temperature fluctuations in the cosmic microwave background (CMB) consistent with a scale-invariant power spectrum. This arises naturally in 'slow-roll' inflationary models from quantum fluctuations of the scalar field which drives the De Sitter phase of exponential expansion, as it evolves towards its minimum.[2] The observed small amplitude, $\delta T/T \sim 10^{-5}$, requires an extremely flat scalar potential stabilized against radiative corrections. This picks out a gauge singlet field in supergravity as the most likely candidate for the 'inflaton'.[4,5]

However such models contain very weakly coupled fields having masses of $\mathcal{O}(M_W)$ and this creates difficulties with the cosmological history *after* inflation. For example, gravitinos can have observable effects on the standard cosmology since they decay very late with lifetime $\sim M_{\rm Pl}^2/m_{3/2}^3$.[6] Although their primordial abundance is inflated away, they are recreated during 'reheating' as the inflaton oscillates about its minimum, converting vacuum energy into radiation. This imposes a severe constraint on the reheat temperature [7] since even a small number of massive late decaying particles can disrupt primordial nucleosynthesis or the thermalization of the CMB.[8] Also inflation offers no solution to the 'Polonyi problem' [9] associated with weakly coupled fields which acquire large vevs along flat directions during the De Sitter phase and release their vacuum energy very late, generating an unacceptable amount of entropy. This is of particular relevance to the moduli in (compactified) string theories.[10]

Despite many ingenious attempts [11] it has proved very difficult to construct a *physical* model of inflation which satisfies these phenomenological constraints, without fine-tuning. Hence the tendency has been to construct rather complicated models with various brand names, viz. 'extended', 'hybrid', 'natural', 'thermal' ...[12] We argue [13] that a simple model based on a singlet field in minimal $N = 1$ supergravity does reach the parts that other, more contrived, models do not reach.

## 2 The Supersymmetric Inflationary Cosmology

The scale of the potential driving inflation is *conservatively* bounded by ascribing the observed CMB anisotropy entirely to gravitational waves; this yields $V^{1/4} \lesssim 4 \times 10^{-3} M_{\rm Pl}$.[14] If the anisotropy is due instead to scalar density fluctuations then the bound is even more restrictive.[15] Thus we can ignore interactions due to string and Kaluza-Klein states with masses of $\mathcal{O}(M_{\rm Pl})$ and use an effective field theory, viz. $N = 1$ supergravity, to describe the inflation sector.

The interaction of gauge singlet fields is then specified by the Kähler potential $K(\Phi^\dagger, \Phi)$ in terms of which the scalar potential is [16]

$$V = \frac{1}{4} e^K [G_a (K^{-1})^a_b G^b - 3 \mid W \mid^2] , \qquad (1)$$

where $G_a = K_a W + W_a$, $W(\Phi)$ is the superpotential, and the indices $a, b$ denote derivatives with respect to the chiral superfields, $\Phi$. Now, apart from the fundamental scale of $M \equiv M_{\rm Pl}/\sqrt{8\pi} \sim 10^{18}\,{\rm GeV}$, the theory contains the gauge symmetry breaking scale $M_{\rm GUT} \sim 10^{16}\,{\rm GeV}$. There must also be a source of supersymmetry breaking characterized by the gravitino mass $m_{3/2} \lesssim 10^3\,{\rm GeV}$, a plausible origin for which is gaugino condensation in a hidden sector; then $m_{3/2} \sim \langle \lambda\lambda \rangle / M^2$ and the gaugino condensate $\langle \lambda\lambda \rangle \sim (10^{13}\,{\rm GeV})^3$. We denote by $\chi, \bar\chi$, the fields which acquire a vev of order $M_{\rm GUT}$ along a $D$-flat direction thus breaking the gauge symmetry, and by $\Theta, \bar\Theta$, the gauge singlet fields with masses of $\mathcal{O}(M)$. Allowing for a coupling between these fields, consider a superpotential of the form (suppressing coupling constants of order unity)

$$W = M\Theta\bar\Theta - \bar\Theta\chi\bar\chi. \qquad (2)$$

We see that the gauge symmetry breaking vev in $\chi, \bar{\chi}$ induces a vev in the massive field: $\langle\Theta\rangle \equiv \Delta = \langle\chi\rangle\langle\bar{\chi}\rangle/M \sim 10^{14}$ GeV. If $\Theta$ couples to other fields, they will acquire mass determined by this vev, according to the strength of coupling. (Similar considerations apply to symmetry breaking in the hidden sector.) Therefore there should be sectors in the theory associated with the mass scale required for acceptable inflation, *without* fine-tuning.

In string theories, which have only one fundamental mass scale, it is probably necessary to drive the vev of $\Theta$ through the SUSY-breaking sector. This is because the vev for the $\chi$ field is induced when the $\chi$ soft SUSY-breaking mass-squared term in the Lagrangian is driven negative by radiative corrections. This cannot occur if there is a large potential energy associated with the inflation sector as this will cause $\chi$ to acquire a soft mass term, inhibiting the development of its vev until inflation is over. Thus the inflationary potential should not exceed the scale of gauge and supersymmetry breaking, otherwise the latter will be inhibited by the very inflationary phase it is supposed to drive. If SUSY-breaking is triggered by gaugino condensation in the hidden sector, the relevant scale is of $\mathcal{O}(10^{13})$ GeV. On the other hand, the requirement of generating sufficiently large density perturbations places a (model-dependent) lower limit on the inflationary scale. We find [13] that both constraints can be satisfied for models which are dynamically of the 'new' [17] rather than the 'chaotic' [18] variety, i.e. in which $\phi$ evolves towards $M$ rather than towards the origin.[a]

The starting point for an inflationary model is the form of the potential describing the inflaton, $\phi$. The coupling between the chiral superfield $\Phi$ (which contains $\phi$ as its scalar component) and $\Theta$ is constrained by the $R$-symmetry of the superpotential (2) under which $\Theta$ and $\chi\bar{\chi}$ transform as $e^{i\gamma}$, $\bar{\Theta}$ transforms as $e^{i(2\beta-\gamma)}$, while the superspace coordinate transforms as $e^{-i\beta}$. The most general superpotential, $P$, describing $\Theta$, $\bar{\Theta}$, $\chi$ and $\Phi$, is then

$$P = \Theta\bar{\Theta}M\, f\left(\frac{\Phi}{M}\right) + \bar{\Theta}\chi\bar{\chi}, \qquad (3)$$

where $f$ is some function which is not constrained by the $R$-symmetry alone. (We have absorbed the constant term generating the $\Theta$ mass in $f$.) As discussed above, once $\chi, \bar{\chi}$ acquire vevs breaking the gauge symmetry, the field $\Theta$ will also acquire a vev $\Delta$ leading to the inflaton superpotential (with $f(0) = 1$)

$$I(\Phi) = \Delta^2 M\, f\left(\frac{\Phi}{M}\right). \qquad (4)$$

---
[a] In order to realise the dynamics of 'chaotic' inflation, one must rely on a small coupling constant (rather than a ratio of mass scales) to provide the scale of inflation in terms of the Planck scale. In string theories small couplings can indeed arise but only dynamically when moduli fields acquire large vevs. These however do not lead to inflationary potentials because the would-be inflationary potential is not sufficiently flat in the moduli direction.[19]

Another way to motivate this form is that it arises due to gaugino condensation in the hidden sector and the inflaton is one of the confined states with mass of $\mathcal{O}(\Delta)$.

We now make the minimal choice of the Kähler potential, $K = \Phi^\dagger\Phi$, corresponding to canonical kinetic energy for the scalar fields. Then the scalar potential following from the superpotential $I$ is [16]

$$V_I(\phi) = e^{|\phi|^2/M^2}\left[\left|\frac{\partial I}{\partial\phi} + \frac{\phi^* I}{M^2}\right|^2 - \frac{3|I|^2}{M^2}\right]_{\Phi=\phi}. \qquad (5)$$

To fix the form of $f(\phi/M)$, we require that supersymmetry remain unbroken in the global minimum, i.e.

$$\left|\frac{\partial I}{\partial\Phi} + \frac{\Phi^* I}{M^2}\right|_{\Phi=\Phi_0} = 0, \qquad (6)$$

and set the present cosmological constant to be zero,

$$V_I(\Phi_0) = 0. \qquad (7)$$

This implies

$$I(\Phi_0) = \frac{\partial I}{\partial\Phi}(\Phi_0) = 0. \qquad (8)$$

The *simplest* form which satisfies these conditions is [5]

$$I = \frac{\Delta^2}{M}(\Phi - \Phi_0)^2. \qquad (9)$$

In order for successful inflation to occur by the 'slow roll' mechanism, the scalar potential must be flat at the origin, $(\partial V_I/\partial\Phi)_{\Phi=0} = 0$ which sets $\Phi_0 = M$. This in turn sets $(\partial^2 V_I/\partial\Phi^2)_{\Phi=0} = 0$ since $I$ does not contain cubic terms. The scalar potential (5) is shown in Fig.1. The imaginary direction is stable while along the real direction we can expand

$$V(\phi) = \Delta^4\left[1 - 4\left(\frac{\phi}{M}\right)^3 + \frac{13}{2}\left(\frac{\phi}{M}\right)^4 + \ldots\right]. \qquad (10)$$

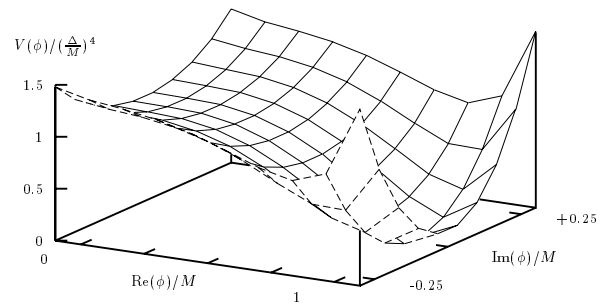

Figure 1: The complex inflationary scalar potential.

Now if the initial value $\phi_i$ is determined by some high temperature effective potential then it is unlikely [20]

that this value coincides with the origin, without fine tuning. However a weakly coupled field which drops out of thermal equilibrium below the Planck scale will, in general, have a broad distribution.[2] While there may be only a small probability that one starts at the point where the first derivative of the potential vanishes, this region *will* inflate and become the overwhelmingly probable state after inflation. In (compactified) string theories, the derivatives of the potential are determined by the vevs of moduli fields. Let us expand $f$ about $\phi_i$,

$$f(\Phi/M) = a(m) + b(m)(\Phi - \Phi_i) + c(m)(\Phi - \Phi_i)^2 + \ldots, \quad (11)$$

in terms of coefficients which depend on the moduli $m$. The coefficient $a(m)$ determines the value of the potential initially so the moduli will flow to minimise this. If the remaining coefficients depend on independent combinations of the moduli they will be undetermined at this stage as they do not affect the initial vacuum energy. Random initial conditions will however allow *some* region in which the value of $b(m)$ is just that needed to make the second derivative of the potential vanish and this will dominate the final state of the universe because of the enhanced amount of inflation it will undergo. Thus most of the features of our inflationary potential are quite natural and to be expected in any theory which yields a potential with a turning point. The exceptional property is that $I$ (9) is a *perfect square*, as is required to ensure the vanishing of the cosmological constant at the minimum. By adjusting the term $c(m)$ we can always arrange that the first three terms form a perfect square, but not in a natural way. This is just a restatement of the usual cosmological constant problem, namely that there is no symmetry which makes it vanish. We rely upon its doing so for unknown reasons, as do all inflationary models.

We now study the dynamics of inflation using the semi-classical equation of motion for the inflaton field [2]

$$\ddot{\phi} + 3H\dot{\phi} + V'(\phi) = 0, \quad (12)$$

where $H \equiv (\dot{a}/a) = \sqrt{V/3M^2}$ is the expansion rate of the cosmological scale-factor $a$ in the vacuum-energy dominated De Sitter phase. The spectrum of scalar density perturbations at horizon-crossing is [15]

$$\delta_H(k) = \sqrt{\frac{1}{75\pi^2}} \frac{1}{M^3} \left(\frac{V^{3/2}}{V'}\right)_\star, \quad (13)$$

where $\star$ denotes the epoch at which a scale of wavenumber $k$ crosses the event horizon $H^{-1}$ during inflation, i.e. when $aH = k$. The CMB anisotropy measured by COBE allows a determination of the perturbation amplitude at the scale corresponding roughly to the size of the presently observable universe, $H_0^{-1} \simeq 3000\, h^{-1}\,\text{Mpc}$, where $h \equiv H_0/100\,\text{km}\,\text{sec}^{-1}\,\text{Mpc}^{-1}$ is the present Hubble parameter. Normalization to the quadrupole moment, $Q_{\text{rms-PS}} \simeq 20\,\mu\text{K}$, in the 2-year COBE data gives $\delta_H \simeq 2.3 \times 10^{-5}$, which fixes the inflationary scale:

$$\frac{\Delta}{M} \simeq 10^{-4}. \quad (14)$$

At the end of inflation, $\phi$ begins to oscillate about its minimum until it decays, reheating the universe. The dominant coupling of $\phi$ to states $\chi$ in another sector with superpotential $P(\chi)$ has the form $(\partial V/\partial \phi)\, P(\chi)_A M^{-2}$ (where the subscript $A$ denotes that the chiral superfields in $P$ should be replaced by their scalar components). This generates a trilinear coupling to the light matter fields $\chi$ of strength $\sim \Delta^2/M^2$, corresponding to a decay width $\Gamma_\phi \sim [m_\phi/(2\pi)^3](\Delta^2/M^2)^2$. With our simplifying assumption that there are no small parameters in $f(\phi)$ the mass of the inflaton is $m_\phi \sim \Delta^2/M$. The inflaton thus reheats the universe to a temperature

$$T_R \sim \left(\frac{30}{\pi^2 g_*}\right)^{1/4} (\Gamma_\phi M)^{1/2} \simeq 10^5\,\text{GeV}, \quad (15)$$

taking $g_* = 915/4$ for the MSSM.[b] This is well below the upper limit imposed by consideration of the production and subsequent decay of unstable gravitinos [7,8]

$$T_R \lesssim 2 \times 10^8,\ 2 \times 10^9,\ 6 \times 10^9\ \text{GeV}$$
$$\text{for}\quad m_{3/2} = 10^2,\quad 10^3,\quad 10^4\ \text{GeV}. \quad (16)$$

Baryogenesis can occur subsequently, e.g. through the late decay of a sfermion condensate.[22]

The direct production of gravitinos from inflaton decay is another potential hazard. The relevant coupling is $h_{\phi\psi_{3/2}\psi_{3/2}} = 2\Delta^2(\phi - M)/M^3$, whereas the coupling to matter fields (dominantly the top squarks and the Higgs in the MSSM) is $h_{\phi^*\tilde{t}\tilde{t}^c H_2} = 2h_t \Delta^2/M^2$. Although both couplings are gravitational in origin, there is a suppression factor $(\phi - M)$ in the gravitino coupling which follows because of the *perfect square* form of the superpotential $I$ (9). This ensures that gravitino production is relatively negligible, $\Gamma_{\psi_{3/2}\psi_{3/2}}/\Gamma_{\tilde{t}\tilde{t}^c H_2} \sim (\Delta/M)^4$.

The 'Polonyi problem' associated with the moduli and dilaton fields is rather subtle and discussed in detail elsewhere.[13] We find only two possible solutions, either that all moduli have vevs fixed by a stage of symmetry breaking *before* inflation, or that the moduli minima are the *same* during and after inflation. The first possibility also requires the dilaton to acquire a mass much higher than the elctroweak scale. In both cases the implication is that the moduli cannot be treated as dynamical variables at the electroweak scale,[23] determining the couplings in the low energy theory.

The detailed implications for large-scale structure formation and CMB anisotropies have been studied elsewhere.[24] Since the potential (10) is dominated by a *cubic* term, the power spectrum of matter density perturbations departs from scale-invariance as $\ln^2(Hk^{-1})_\star$, giving a 'tilted' spectrum with a slope of about 0.9. As shown in Fig.2,[24] this reduces the power on galactic scales in a cold dark matter (CDM) cosmogony,[25] thus providing a better match than the usually assumed scale-invariant spectrum to the 'data' inferred from observations of large-scale structure. The cosmological parameters used here are $\Omega = 1$, $\Omega_N h^2 = 0.0125$, $h = 0.5$; an improved fit can be obtained by lowering $h$ further.[26]

---

[b]'Parametric resonance'[21] does not occur since the inflaton has no coupling of the form $\phi^2\chi^2$ but only terms involving $\chi^3$.

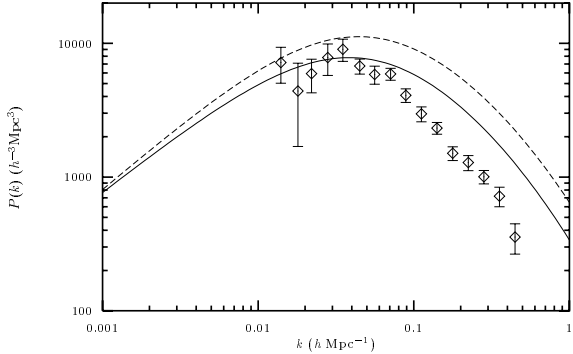

Figure 2: The COBE-normalized power spectrum of density fluctuations in cold dark matter (solid line) compared with *inferred* data. The dashed line shows the 'standard' CDM model.

Another observational probe is the power spectrum of the CMB angular anisotropy.[27] A complication here is that primordial gravitational waves also contribute to the quadrupole moment measured by COBE, but not to the higher multipoles being probed presently by experiments measuring anisotropy on small angular scales. However in our model, the generation of gravitational waves is negligible because the inflationary potential is so very flat. Numerical solution of the coupled Boltzmann equations for the radiation and matter fluids then yields the angular power spectrum shown in Fig.3.[24] More accurate observations will permit a definitive test.

## Acknowledgments

I thank Jenni Adams and Graham Ross for enjoyable collaborations and for allowing me to present our results. My work is supported by a PPARC Advanced Fellowship.

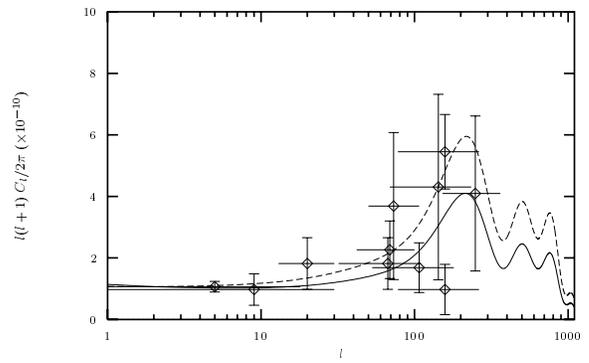

Figure 3: The COBE-normalized power spectrum of CMB temperature fluctuations (solid line) compared with observations. The dashed line shows the 'standard' CDM model.